# PUSHING STEM-EDUCATION THROUGH A SOCIAL-MEDIA-BASED CONTEST FORMAT – EXPERIENCES AND LESSONS-LEARNED FROM THE H2020-PROJECT SCICHALLENGE

**Florian Huber[1], Bernhard Jäger[1], Sabri Pllana[2], Zdenek Hrdlicka[3], Christos Mettouris[4], George A. Papadopoulos[5], Tamara Matevc[5], Zsófia Ocsovszky[6], Edina Hajdu[6], Chris Gary[7], Phil Smith[8]**

[1] *SYNYO GmbH (AUSTRIA)*
[2] *Linnaeus University (SWEDEN)*
[3] *University of Chemistry and Technology (CZECH REPUBLIC)*
[4] *University of Cyprus (CYPRUS)*
[5] *Jožef Stefan Institute (SLOVENIA)*
[6] *BioTalentum Ltd (HUNGARY)*
[7] *Kinderbüro Universität Wien (VIENNA)*
[8] *Teacher Scientist Network (UNITED KINGDOM)*

## Abstract

Science education is a crucial issue with long-term impacts for Europe as the low enrolment rates in the STEM-fields, including (natural) science, technology, engineering and mathematics, will lead to a workforce problem in research and development. In order to address this challenge, the EU-funded research project SciChallenge (project.scichallenge.eu) aims to find a new way for getting young people more interested in STEM. For this purpose, the project developed and implemented a social-media-based STEM-contest for young people, which aims at increasing the attractiveness of science education and careers among young people.

In the first two parts, the paper reflects on the problem, introduces the project and highlights the main steps of the preparation of the contest. The third section of the paper presents the idea, design and implementation of the digital contest platform (www.scichallenge.eu), which serves as the core of the challenge. The fourth part of the paper will provide a status update on the contest pilot. It will provide a summary of the experiences that the consortium made with this novel approach as well as the main obstacles that the consortium was facing. The paper will conclude with a preliminary reflection on the question if such an approach can help to increase the interest of young people in STEM-education and careers.

## 1 STEM EDUCATION IN EUROPE

Due to the global division of labour, the outsourcing of industrial production to other parts of the world and the rise of a knowledge-based economy, scientific and technological innovations have become increasingly important. The economies and societies of the EU-member-states are very much dependent on the research and development capacities. Hence, Europe's research and development efforts form an integral part of the European economy.

The achievements in science and technology have been significant and scientific research in Europe is supported or conducted by industry, universities and scientific institutions. Europe is home to some of the world's oldest universities, such as the University of Bologna or the University of Oxford. Moreover, excellent science and research has been Europe's key asset for sustainable growth for many years and it has maintained the leading position in a highly competitive global economy. National governments, stakeholders, in particular research funding and performing organisations, as well as European Institutions are pooling their resources in order to increase the efficiency and impact of research and to co-operate across borders to allow researchers and knowledge to flow freely within Europe. [1]

According to the National Science Foundation, societies that want to succeed in this new era need a strong research sector and hence they need to promote research careers far more than in the past. [2]

Science and research are determining the economic prosperity in the modern global economy as well as the global competitiveness.

Nevertheless, the enrolment rates in natural science, technology, engineering and math (STEM) still lack behind other disciplines. This will create a workforce problem in the future. The 2009 report "STEM Supply and Demand Research" additionally states that numbers are even sugar-coated: The decline of students in the fields of mathematics, engineering and physics, is hidden due to the growth of students in IT and the biology. [3]

This holds especially true for the Central and Eastern European Countries (CEE), which were and are still facing a massive decline in the "classical" industrial production and are therefore more and more dependent on the growth of the high-tech and knowledge industries. In order to stay competitive on the European and global market, they need to increase their research and development capacity, which can only be facilitated by higher numbers of researchers and a general increase of the research sector. [3]

A common problem in many countries is the fact that the broad variety of career options in science and research is not visible. Many students are only focussing on positions at public universities. However, these institutions are often affected by austerity measures and hence the positions in this sector are decreasing. Therefore, young researchers need to be aware and take into account the many other options that private research organisations and SMEs offer.

In order to engage the interest in science and research careers also beyond the rather classical options, it is therefore important to establish a link between the universities and applied research organisations or research-intense industries to overcome this problem. Such a new angle would open new pathways for future careers in science and research.

Science Europe, an association of European Research Funding Organisations (RFO) and Research Performing Organisations (RPO), is actively involved in the development of the European Research Area (ERA). Its policy-related work is guided by the Science Europe Roadmap and it outlines the strategic objectives for Science Europe. Furthermore, Science Europe defined nine priority action areas – with one of them being "Research Careers". [4]

The topic of "research careers" encompasses not only the awareness for the different strands of careers that are possible, but also the conditions under which researchers pursue their professional endeavour. The lack of options for woman and minorities to compete for employment, wages, and leadership in science was already shown in 1990. [5] And although many efforts have already been made to overcome structural inequalities, we are still far away from equality. [6] Hence, the career conditions need to include optimized recruitment processes, career progression at all levels, and equality regardless of gender, ethnicity or other differences.

In order to ensure the vitality of the European research system, new generations of talented people must be attracted into research professions by promoting the broad variety of the STEM-disciplines and by highlighting the many different career options in research. Young boys and girls therefore have to be engaged to pursue careers in Science, Technology, Engineering and Mathematics (STEM). But how can we manage it to get young people more interested in STEM?

## 2   THE SCICHALLENGE CONTEST

The EU-funded research project SciChallenge (project.scichallenge.eu) addresses this challenge by proposing a social-media-based STEM-contest for young people between 10 to 20 years. The contest pilot is currently running (until April 30th 2017). With its multi-level approach, SciChallenge aims at increasing the attractiveness of science education and careers among young girls and boys on a pan-European level. Furthermore, several awareness creation modules as well as a strong social-media presence will be implemented in order to make STEM education and careers more interesting.

As Fig. 1 shows, the SciChallenge project is follows a "waterfall" concept and it is structured along several tasks. The focus of the first task is to collect knowledge, stakeholders and good practices as well as science contests, which are analysed in order to provide inputs for the SciChallenge contest development. The next task is dedicated to the contest preparation and includes the contest conceptualization as well as the creation of contest guidelines and the preparation of STEM-topics as an inspiration for the participants. The following task is focussing on the development of a digital platform, which serves as the core of the SciChallenge contest. Next, the contest is implemented and promoted through different channels with a particular focus on social media. Strongly related with this work, the following task is focussing on creating awareness for STEM-careers and in particular for research-based companies, which offer internships and taster days. The final task is dedicated to the dissemination, communication and distribution of results, that is also a continuous activity throughout the entire project.

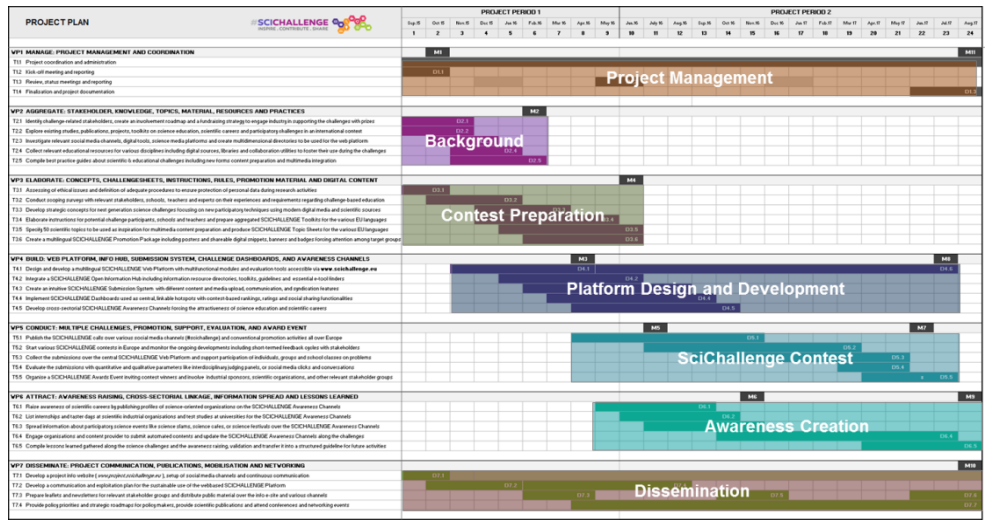

**Fig. 1 Project Concept**

As it was highlighted above, one major outcome of work package 3 was the development of an entirely novel concept for setting up the SciChallenge contest. During the development of the contest framework, several aspects such as eligibility, the topics, the rating and awarding as well as the beneficial integration of social media needed to be taken into account. The next section will highlight, how the SciChallenge consortium set up the contest framework.

## 2.1 Processual Contest Framework

The SciChallenge contest infrastructure basically consists of two main and intertwined parts: On the one hand, it includes the processual contest framework, which describes how to set-up and implement this social-media based contest. On the other hand, it includes the digital contest platform (see Section 3), which serves as the core instrument for running the contest. [7]

The processual contest framework includes four main phases, which are explained below and visualized in Fig. 2:

(1) The first phase of the SciChallenge contest focusses on the eligibility. The SciChallenge contest is conceptualized as a pan-European contest. Hence, the country of residence of the participants has to be a EU-member-state. Regarding the age groups, the contest aims at participants between 10 to 20 years. In order to ensure a fair contest, the age scope is divided into two age groups, with the 1st age group ranging from 10 to 14 years and the 2nd age group ranging from 15 to 20 years. Furthermore, other criteria might be included optionally included – based on the decisions of the consortium during the preparation.

(2) Eligible participants then enter the second phase of the contest, the participation. SciChallenge allows both for individual or group participation. After this is decided by the participants, the participants select a specific topic based on 50 topic sheets, which are provided by the consortium as an inspiration, and develop a project on this issue. The contribution for the contest can be done either as a poster, a presentation or as a video (additional media formats might be included by the consortium).

(3) After the participants have finalized their contribution for the contest, they enter the third phase of the contest concept: the submission. In a first step, the participants upload their contribution to one of the pre-defined social media sites and include the hashtag #SciChallenge2017. In a second step, the participants register on the digital contest platform and place the link to their contribution in the submission system. The contributions will be stored in a database and after the submission portal closes, all contributions will be aggregated.

(4) The final phase of the SciChallenge contest focusses on the awarding process: (a) The first step of the rating procedure is conceptualized as an online community rating and derives from the social media sites, where the participants uploaded their contribution. An automated system integrated in the SciChallenge platform will use the submission links and the hashtag #SciChallenge for counting the number of "views", "likes" and "shares" of every contribution. Based on this online community rating, a first ranking of the best contributions is assembled. (b) The second step of the rating procedure is conceptualized as a jury rating, where a jury will rate those contributions that made it through the online community rating. Of course the jury will vary regarding size and expertise depending on the topic and the purpose of the contest. Independent from that, the jury members can access an intuitive rating module, where they will rate the contributions that were successful in the online community rating. (c) Based on the jury rating, 12 winners of the SciChallenge contest will be selected and they will be invited to the final award event in Vienna, where the winners present their project, receive the jury award as well as an additional audience award.

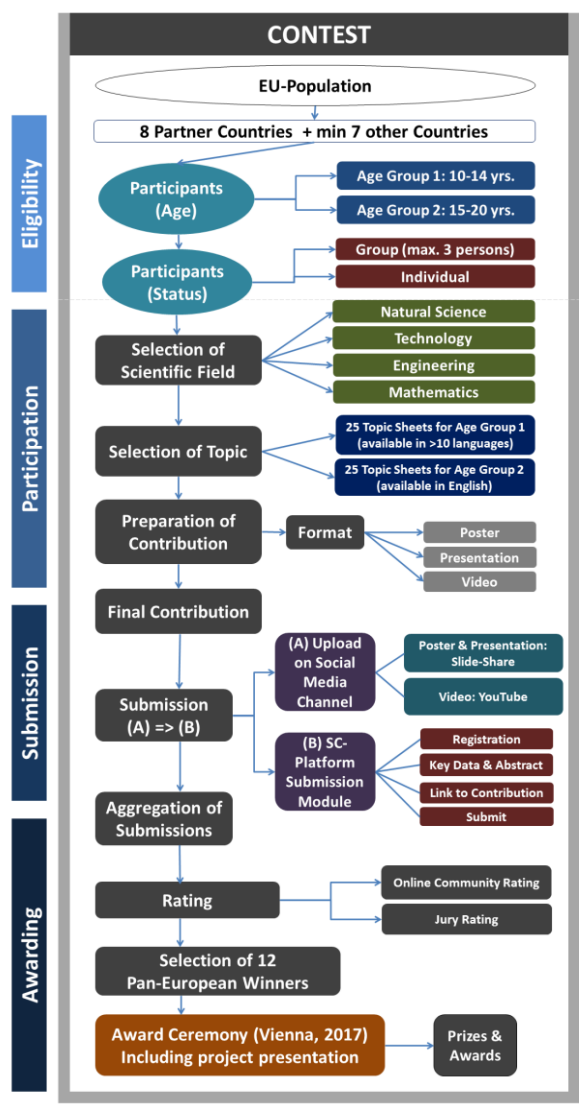

**Fig. 2 SciChallenge Processual Contest Framework**

## 2.2 Provided Contest Resources

Generally, there are two major approaches to definition of tasks in the existing STEM contests: (1) precisely defined problems or questions that contestants have to solve or answer, (2) broad categories or topics that contestants can choose to develop their projects.

In order to keep the SciChallenge contest as open as possible and to encourage creative approaches to a particular topic, the consortium followed the second approach. For this purpose, a set of SciChallenge topic sheets was developed, which should inspire potential participants to take part in the SciChallenge contest. Hence, the topic sheets serve as a catalyst for the contestants to start working on their contributions.

A SciChallenge topic sheet conceptually is a single-page document, which provides the context about a particular topic, related societal challenges, several keywords to search for further information, and an inspiring image or illustration. The intention of topic sheets is to help participants to start reasoning about the topic. Fig. 3 shows the topics provided in the topic sheets.

| ID | Topic Title | ID | Topic Title |
|---|---|---|---|
| AG1_01 | Food Chemistry | AG2_01 | Online Life |
| AG1_02 | Health | AG2_02 | Food Production |
| AG1_03 | Biotope: Life in Water | AG2_03 | Ageing |
| AG1_04 | Cloning | AG2_04 | Prenatal Development |
| AG1_05 | Minerals/Geology | AG2_05 | Stem Cell |
| AG1_06 | Colour | AG2_06 | Oil Peak |
| AG1_07 | Dirt | AG2_07 | Clean Water |
| AG1_08 | Smoking | AG2_08 | Climate Change |
| AG1_09 | Biodiversity | AG2_09 | Human Migration and Evolution |
| AG1_10 | Natural Hazards | AG2_10 | Assistive Technologies |
| AG1_11 | Cyber Bullying | AG2_11 | Big Data |
| AG1_12 | Search Engines | AG2_12 | Internet of Things |
| AG1_13 | Social Media | AG2_13 | Cybersecurity |
| AG1_14 | Internet | AG2_14 | Active Assisted Living |
| AG1_15 | Agricultural Engineering | AG2_15 | Nano-Bio-Technology |
| AG1_16 | Mobility | AG2_16 | Advanced Materials |
| AG1_17 | Robotics | AG2_17 | Energy Efficiency |
| AG1_18 | Future Cities | AG2_18 | Food Transportation |
| AG1_19 | Food Preparation | AG2_19 | 3D Printing |
| AG1_20 | Solar Power | AG2_20 | Food Recycling |
| AG1_21 | Gravity | AG2_21 | Green Buildings |
| AG1_22 | Bionics | AG2_22 | Crop Disease |
| AG1_23 | Nanotechnology | AG2_23 | High Frequency Trading |
| AG1_24 | Golden Ratio | AG2_24 | E-Health |
| AG1_25 | Cryptography | AG2_25 | Cyber Scams |
| AGX_51 | Open STEM Topic | AGX_51 | Open STEM Topic |

**Fig. 3 List of Topics for AG1 and AG2**

The consortium has developed 50 topic sheets that define specific topics along the four STEM disciplines: natural science, technology, engineering, or mathematics. Additionally, an open topic sheet (#51) enables contestants to participate with a STEM-topic of their choice. Basically the topic sheets target pre-university students between 10 and 20 years. However, 25 topic sheets were developed for the younger age group (10-14 years) and these topic sheets were also translated into 10 languages. The other 25 topic sheets are targeting the older age groups (15-20 years) and are available only in English.

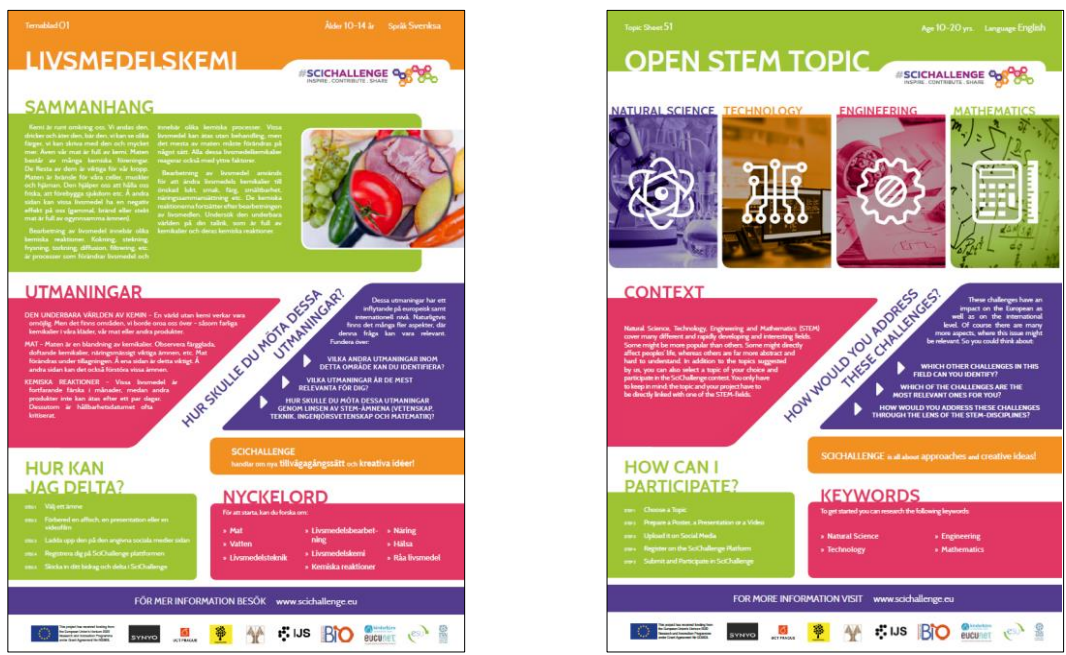

**Fig. 4 Topic sheet in Swedish and Open STEM Topic in English**

In order to make the process of applying to the SciChallenge contest as simple and understandable as possible for potentially interested participants of all age groups, two Scichallenge Toolkits (Fig. 5) – in the sense of instructions or guidelines – were prepared.

The first toolkit aims at the participants or those who intend to prepare a contribution for the contest. It provides more detailed information on the eligibility criteria as well as on the formats that need to be used. The detailed specifications as well as the step-by-step guide on how to participate should make it rather easy for the students to be part of the contest.

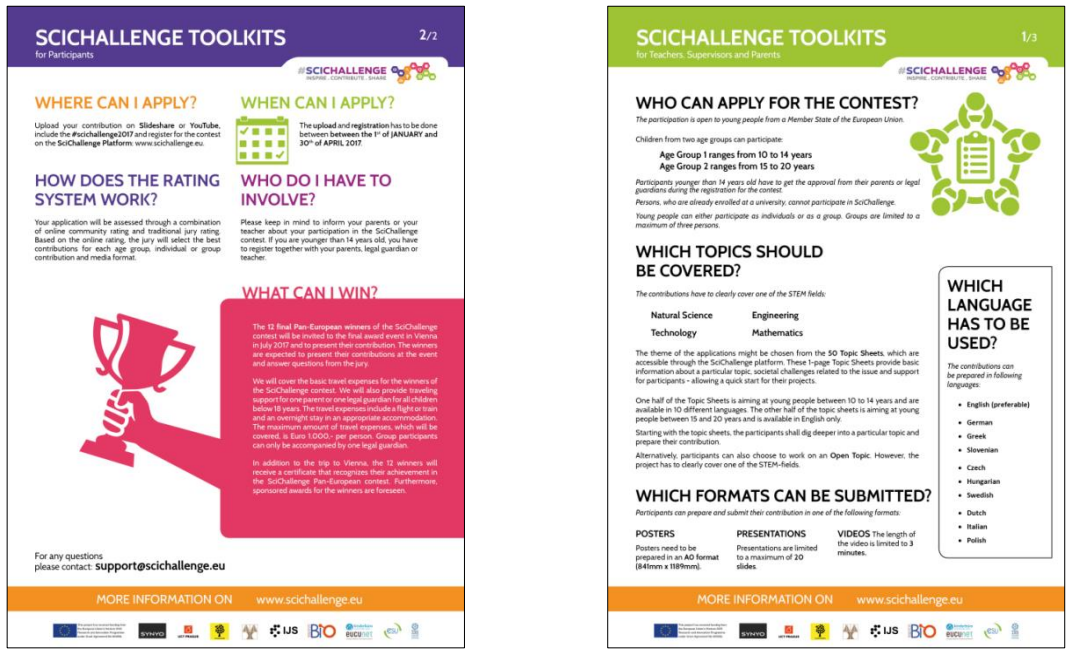

**Fig. 5 Examples from the toolkits for participants and supervisors**

The second toolkit aims at teachers and parents, who are guiding participants through the preparation and submission phase. Hence, it provides more detailed information on the participation and on further criteria as well as on the rating and awarding.

## 2.3 Rating Calculation

The rating of the SciChallenge is divided into both an (A) online community rating and (B) a jury rating. As it is defined in the submission system, each contribution falls into one of the twelve categories, which reflect the two age groups and the (for each age group) six media types. Fig. 6 highlights the rating calculation.

In the (A) online community rating, the automated rating system collects the views and likes of each contribution. Based on the meta-data from the submission system, it furthermore groups the contributions along the twelve categories as well as along the country of residence of the participants.

The top submission of each participating country (target: minimum of 15 countries) and of each category (12 categories) will then proceed to (B) the jury rating. The jury, consisting of educational experts from different countries, will rate the contributions along several criteria by using a scoring matrix. The criteria include: presentation of the problem, creativity of the realization, added value, future thinking and (for the older age group) scientific approach.

Based on the jury rating, the 12 winners of the SciChallenge contest will be selected and awarded at the award event in Vienna in July 2017.

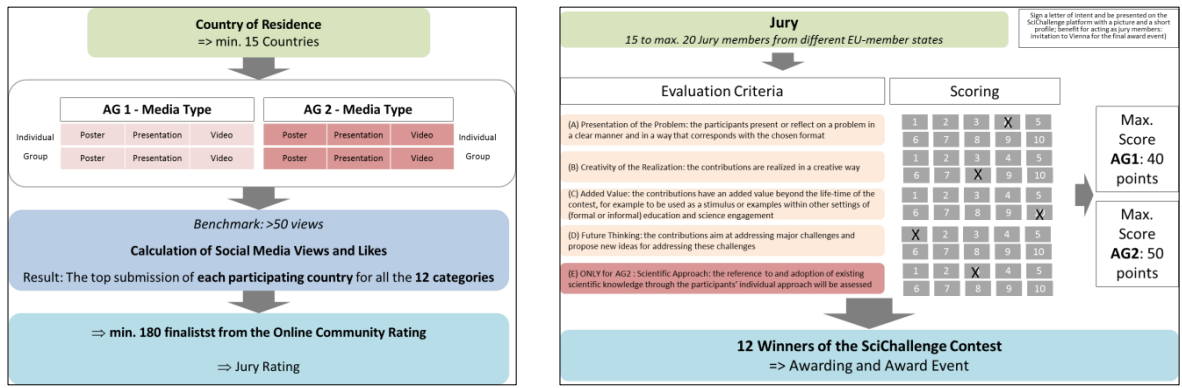

**Fig. 6 Rating calculation**

## 3 DIGITAL CONTEST PLATFORM (WWW.SCICHALLENGE.EU)

### 3.1 Platform Concept

The SciChallenge platform is basically structured along the main dimension of the SciChallenge Processual Contest Framework presented in Fig. 2.

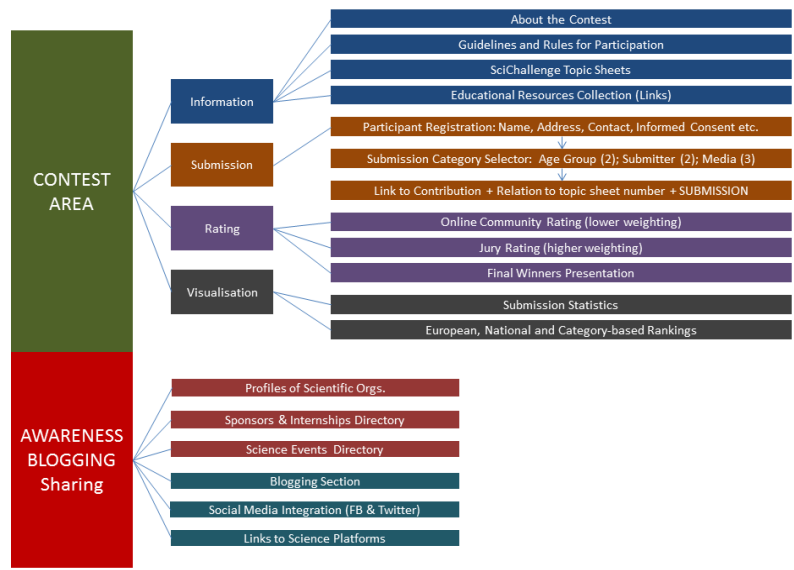

**Fig. 7 SciChallenge Platform Concept**

As Fig. 7 shows, it includes several (A) Contest Modules, which aim at providing information, allowing the submission of the contributions, calculating the rating and visualizing the ongoing contest activities. Furthermore, several modules are dedicated to (B) Awareness Creation, Blogging and Sharing.

### 3.2 Platform Modules and Functionalities

The homepage is the entry point of the SciChallenge platform and offers quick access to all relevant resources for the participants. An animated header shows the steps for a successful participation in the contest. The contest board allows direct access to the registration and submission portal as well as to the toolkits and the topic sheets.

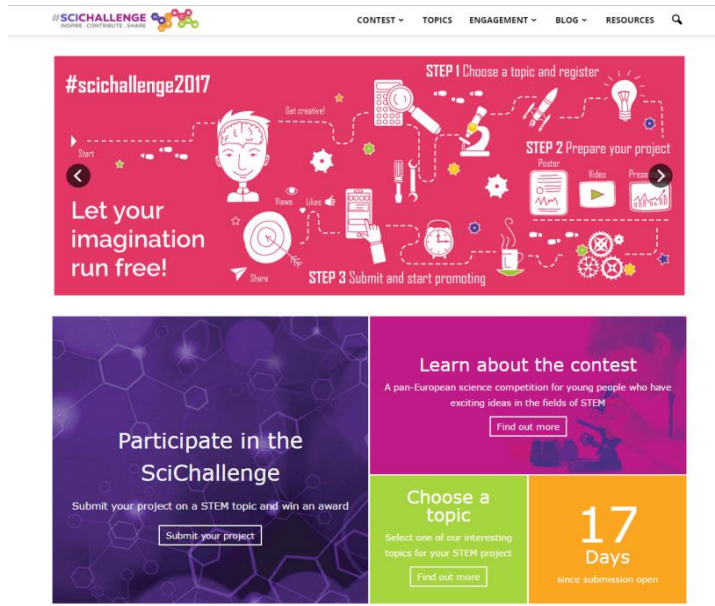

**Fig. 8 Entry page of the platform**

Once the participants prepared their contribution for the SciChallenge contest and are planning to submit it, they need to create an account in the SciChallenge submission system. The first step of registration requires putting in the first name, the last name, an e-mail address and a password. After the registration, they are asked to provide basic information for the contest in the "Profile" section.

In the next step, the participants describe their project and select the topic sheet to which their project is related to. Then, they select the type of format, which they used for their project and finally, they put in the direct link to their contribution on Slideshare or YouTube. Finally, the participant can review and finally submit the contribution.

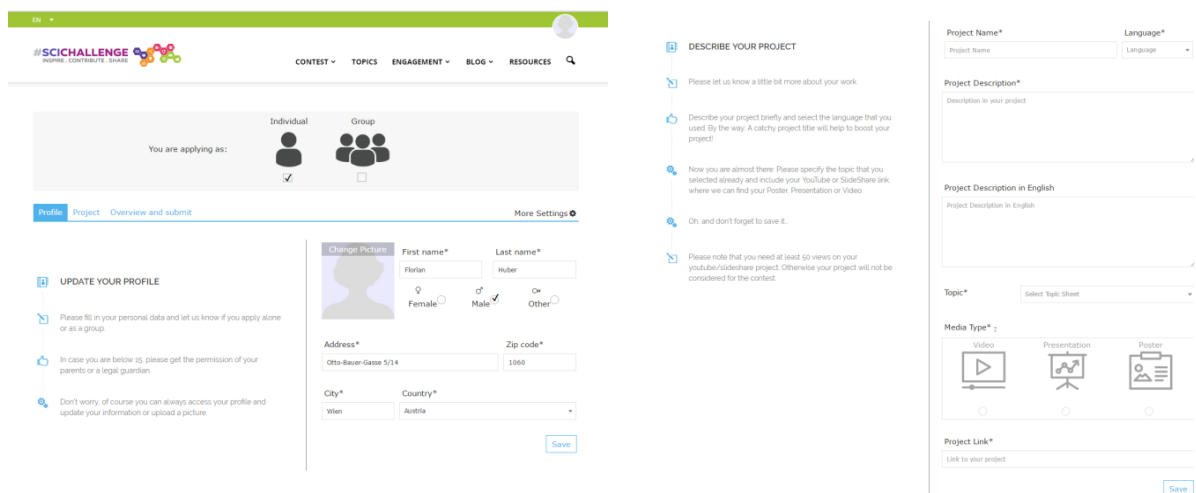

**Fig. 9 Exemplary views of the Contest Modules**

The aim of the awareness channels is to present engagement opportunities to the users of the platform and especially the participants in the SciChallenge contest. These "Engagement" channels that are implemented on the platform include profiles of research-based companies and scientific organizations in order to show the variety of fields and career options in the STEM-fields, internships opportunities, open days and taster days at universities and science events directories on the national level.

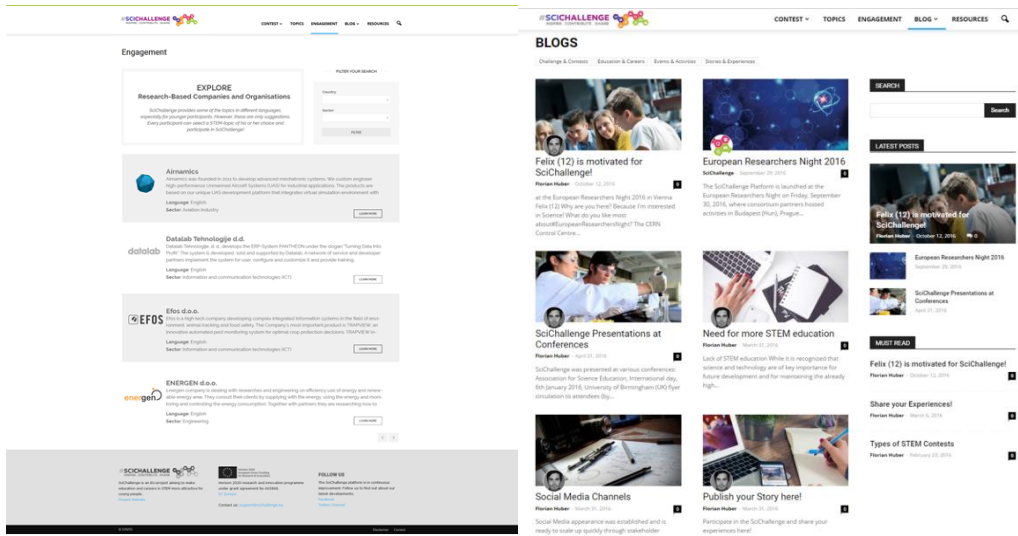

**Fig. 10 Exemplary views of the Awareness Modules**

## 3.3 Social Media Syndication

A specific innovative feature that will also be included in the SciChallenge platform is the automated social media (web) syndication consisting of different tools, which aim at both getting content from the contest disseminated and bringing potential participants and target groups to the platform.

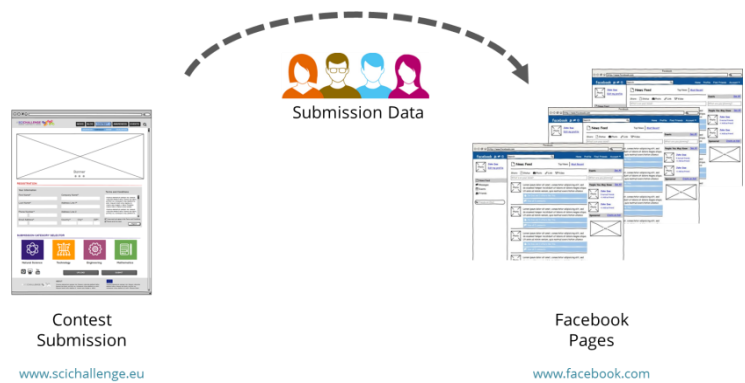

**Fig. 11 Social Media Syndication**

On the one hand, plug-ins will create posts on different social media sites, for example when an applicant submits a contribution or when content on the awareness channels is updated. On the other hand, participants can also create their own personalized widgets related with their contribution and they can then post the widget on their social media page. This will direct new potential participants to the platform and to the awareness channels. The benefit: the participants will take over a major part of the contest promotion by sharing the information on social media.

## 4 EXPERIENCES AND LESSONS-LEARNED

At the time of the submission of this paper (January 19, 2017), the SciChallenge contest – being open for submission from January 1 to April 30 – was still at a very early stage. However, we can already draw initial conclusions regarding the main obstacles the consortium was facing, which might be useful for further initiatives and contests on STEM and education in general.

One of the major lessons-learned of the project is that some of the promises of social media did not work as it was expected. Social media for example do not redeem you from traditional contacting and networking activities, which also require – at least to some extent – physical presence. Relying only on social-media for promoting the SciChallenge contest for example did not create the expected results as it is very difficult to reach the target group (young people between 10 to 20 years) only through social media.

Hence, the consortium soon developed and applied a cross-channel and cross-stakeholder strategy for promoting the STEM-contest:

A) *Social Media*: after 10 months of generally building up the main social media channels Facebook and Twitter and after feeding them with similar content, the consortium decided to clearly separate these channels. Since then, Facebook is mainly targeting potential participants of the contest (as well as facilitators such as teachers) and Twitter is mainly reaching out to networks, scholars and educational professionals (as well as to facilitators).
B) *Traditional Media*: Still traditional media plays a crucial role for getting public attention. Hence, several consortium partners used their professional contacts to organize interviews and media reports. Furthermore, media cooperation was established at least in some of the partner countries, which also aims at reaching potential facilitators and multipliers.
C) *Educational Administration*: It already became apparent in the expert interviews, which the SciChallenge consortium conducted at the beginning of the project that it is crucial to seek for support of the educational administration (such as Ministries of Education) and invite them to promote the contest officially. Hence, the consortium had appointments and contact with representatives from the education administration, which supported the contest through their channels. These activities helped to get recognition for the contest especially among schools and teachers, who then perceived the contest as being sort of an "official" activity.
D) *Initiatives and Networks*: There are several European networks of education experts and teachers and the consortium approached them through newsletters as well as through direct contacting by consortium partners. These networks turned out to be crucial partners for the promotion. For example, when Scientix and the STEM-Alliance both tweeted about SciChallenge, the traffic on both the SciChallenge platform and the Twitter channel multiplied.
E) *Educational Fairs*: In addition to the digital and stakeholder channels, educational fairs and dedicated European events such as the European Researchers Night 2016 (ERN2016) turned out to be perfect for getting directly in contact with young people. For example, the consortium used the ERN2016 as a date for an initial launch of the SciChallenge Platform. Several consortium partners in different countries were present at the ERN2016 and – among other activities – they presented the contest and the platform to young people. Of course the consortium is aware that mainly young people, who are already interested in science, are attending such events. However, they can also be seen as multipliers for encouraging other classmates and friends to participate in the contest or generally in educational activities.

Based on the experiences from the SciChallenge project we can therefore conclude that a social media based approach such as the SciChallenge contest is capable of getting young people interested in STEM or education in general. However, it is crucial to (1) clearly define the focus of each social media channels that is used, and to (2) integrate social media activities into a broader multi-dimensional outreach strategy. This strategy need to include key players from the educational system as well as multipliers from different networks and it also needs to combine digital and non-digital communication channels in order to get the target audience on board.

## ACKNOWLEDGEMENTS

The project SciChallenge has received funding from the European Union's Horizon 2020 Research and Innovation Programme under Grant Agreement No 665868.